\documentclass[12pt]{article}
\begin{document}
\title{{\bf Do Our Observations Depend
upon the Quantum State of the Universe?}
\thanks{Alberta-Thy-13-09, arXiv:0907.4751}}
\author{
Don N. Page
\thanks{Internet address:
don@phys.ualberta.ca}
\\
Theoretical Physics Institute\\
Department of Physics, University of Alberta\\
Room 238 CEB, 11322 -- 89 Avenue\\
Edmonton, Alberta, Canada T6G 2G7
}
\date{(2009 July 30)}

\maketitle
\large
\begin{abstract}
\baselineskip 18 pt

Generically the probabilities of observational results depend upon both
the quantum state and the rules for extracting the probabilities from
it.  It is often argued that inflation may make our observations
independent of the quantum state.  In a framework in which one considers
the state and the rules as logically separate, it is shown how it is
possible that the probabilities are indeed independent of the state, but
the rules for achieving this seem somewhat implausible.

\end{abstract}
\normalsize

\baselineskip 13.9 pt

\newpage

\section*{Introduction}

A goal of science is to produce complete theories $T_i$ that each
predict normalized probabilities $P_j(i)$ of observations $O_j$,
\begin{equation}
P_j(i) \equiv P(O_j|T_i)\ \mathrm{with}\  \sum_j P_j(i) = 1.
\label{prob}
\end{equation}

Assuming that a complete physical theory of the universe is quantum, I
would argue \cite{BA} that it should contain at least the following
elements:\\
(1) Kinematic variables (wavefunction arguments)\\
(2) Dynamical laws (`Theory of Everything' or TOE)\\
(3) Boundary conditions (specific quantum state)\\
(4) Specification of what has probabilities\\
(5) Probability rules (analogue of Born's rule)\\
(6) Specification of what the probabilities mean\\

Here I shall call elements (1)-(3) the quantum state (or the ``state''),
since they give the quantum state of the universe that obeys the
dynamical laws and is written in terms of the kinematic variables, and I
shall call elements (4)-(6) the probability rules (or the ``rules''),
since they specify what it is that has probabilities (here taken to be
the results of observations, $O_j$, or ``observations'' for short), the
rules for extracting these observational probabilities from the quantum
state, and the meaning of the probabilities.  What I shall write below
is largely independent of the meaning of the probabilities, though
personally I view them in a rather Everettian way as objective measures
for the set of observations with positive probabilities. 

Usually it is implicitly believed that the observational probabilities
depend strongly upon the quantum state.  (Sometimes the Everett
interpretation \cite{Everett} is taken to mean that all of physical
reality is determined purely by the quantum state, without the need for
any additional rules to extract probabilities, but this extreme view
seems untenable \cite{Kent} and will not be adopted here.  Instead, I
shall discuss the opposite view, that the probabilities are independent
of the quantum state.)  However, some advocates of inflation\cite{NS,
AS,GSVW,ELM,SPV,Vil06,VV,Vil07,Linde06,Win06,Vanchurin,Linde08b,LWb,
Win08a,SGSV,GV09,LVW,Albrecht} often claim that our observations do not
depend upon the quantum state at all, but rather that inflation acts as
an attractor to give the same statistical distribution of observations
from any state.

In this note, I shall use the framework of state plus rules to discuss
this possibility that observational probabilities might be independent
of the quantum state.  I shall show that this indeed is logically
possible, but apparently only if the probability rules are rather {\it
ad hoc}.  If indeed the rules are this {\it ad hoc}, so that the
probabilities of our observations do not depend upon a quantum state at
all, it would seem to leave it mysterious why many of our observations
can be simply interpreted as if our universe really were quantum.

\section{States and rules}

Let me first discuss the logical independence of the quantum state and
the probability rules.  I shall assume that even if one fixes the
kinematic variables and the dynamical laws, there remains freedom in
what the quantum state is (e.g., many different solutions to the same
Schr\"{o}dinger equation with the same arguments and the same
Hamiltonian, or many different solutions to the constraint equations of
quantum gravity).  The set of all quantum states obeying whatever
kinematic and dynamical constraints one might impose I shall call the
state space; it might or might not be a Hilbert space.  The states
themselves might be pure states, density matrices, or C*-algebra states,
but I shall assume that they are at least C*-algebra states, so that
each state gives the expectation value of the kinematically allowed
quantum operators.  For simplicity, I shall often assume that the
quantum state is a pure state in a finite-dimensional Hilbert space,
though most of the discussion should be generalizable to any C*-algebra
states.

In traditional quantum theory, observational results are eigenvalues of
a certain self-adjoint operator called an observable, which in the
finite-dimensional case at least can be written as a sum of orthonormal
projection operators (formed from the eigenstates of the observable, or
from the eigenspaces of eigenstates for degenerate eigenvalues)
multiplied by coefficients that are the eigenvalues of the observable.
Then the observational probability of each eigenvalue is given by Born's
rule \cite{Born} as the expectation value of the corresponding
projection operator in the quantum state of the system.  In this case,
the logical freedom of the probability rules is the freedom to choose
the observable whose eigenvalues represent the observational results.

In the case of a pure state in a finite-dimensional Hilbert space, the
state by itself does not determine the observational probabilities,
since the probabilities also depend upon the orthonormal projection
operators corresponding to the observable.  Furthermore, any other pure
state in the same Hilbert space would give the same probabilities for
another observable obtained simply by transforming the original
observable by the same unitary transformation used to transform the
state from the original one to the final one.  (This unitary
transformation is not uniquely defined, since only its action on the
original quantum state is specified, so there is a large set of
different transformed observables that all give the same probabilities
as well.)  Then the probability of an eigenvalue of the new observable
in the new state would be the same as that of the eigenstate of the
originaly observable in the original state.  Thus states by themselves
do not determine probabilities, and all pure states give the same
probabilities when they are paired with corresponding observables.  It
is only the relation between the state and the observable that
determines unique probabilities by Born's rule.

In cosmology with a universe large enough that there may be copies of an
observer, no matter how precisely it is described locally, Born's rule
does not work \cite{cmwvw,insuff,brd,BA} and must be replaced by another
set of rules for extracting observational probabilities from the quantum
state.  Generically then the ambiguity in the rules is even greater than
in traditional quantum theory in which one needed to specify just one
quantum operator, a single observable, in addition to the state.

Here for simplicity I shall focus on cases in which the set of rules are
that observational probabilities are obtained by normalizing a set of
unnormalized measures that are each given by the expectation value of a
positive operator in the quantum state,
\begin{equation}
P_j(i) = \frac{p_j(i)}{\sum_{k}p_k}\ \mathrm{with}\  
p_j(i) = \langle \mathbf{q}_j \rangle_i,
\label{normalized-probabilities}
\end{equation}
where $\mathbf{q}_j$ is the positive operator corresponding to the
observational result $O_j$ (or observation $j$, for short), and where
$\langle\ldots\rangle_i$ denotes the quantum expectation value, of
whatever operator replaces the $\ldots$ inside the angular brackets, in
the quantum state $i$ given by the theory $T_i$.  Then, instead of the
single observable required to give the probability rule in traditional
quantum theory by Born's rule, one needs a whole set of positive
operators $\mathbf{q}_j$, one for each observation $j$.

Quantum theories of this form may be axiomatized by the following two
axioms:

{\bf State}:  {\it There is a quantum state that gives
expectation values of operators.}

{\bf Rules}:  {\it Each possible observation has a corresponding
positive operator whose expectation value in the quantum state is the
measure for that observation.}

When we want to do a Bayesian analysis and compare different theories
$T_i$ for which we have assigned prior probabilities $P(T_i)$, we would
like to normalize the measures for observations by dividing by the total
measure and then interpret the normalized measures as the likelihoods or
the probabilities of the observation given the theory, $P(O_j|T_i)$. 
Then if we had a complete set of theories for which we assigned nonzero
prior probabilities, so $\sum_i P(T_i) = 1$, then the posterior
probability of theory $T_i$, given the observation $O_j$, would be given
by Bayes' formula as
\begin{equation}
P(T_i|O_j) = \frac{P(T_i)P(O_j|T_i)}{\sum_l P(T_l)P(O_j|T_l)}.
\label{post}
\end{equation}

Under the assumption that observations are conscious perceptions, the
operators $\mathbf{q}_j$ whose expectation values would then give the
measures of the corresponding conscious perceptions were called {\it
awareness operators} in my previous work \cite{SQM,pim,MS,Page-in-Carr},
but in \cite{cmwvw,insuff,brd,BA} and here I am not restricting to the
assumption that observations must be conscious perceptions (though I
have not given up my personal belief that the most fundamental
observations are indeed conscious perceptions).  The only restriction on
observations I am making here is that each of them should be a complete
observation in the sense that no observation is a proper subset of
another observation.  Here, let us call the $\mathbf{q}_j$ {\it
observation operators}, since it is their expectation values that give
the ratios of observational probabilities.

\section{Rules giving state-independent probabilities}

Now let us consider whether we can have probability rules giving
observational probabilities independent of the quantum state, as is
often claimed or wished to be the case for inflationary universes.  It
is clear that if $\langle \mathbf{q}_j \rangle_i$ is to be independent
of the quantum state, the observation operator $\mathbf{q}_j$ must be
proportional to the identity operator, $\mathbf{q}_j =
p_j(i)\mathbf{I}$, with each nonnegative $p_j(i)$ that can be chosen
arbitrarily and independently of the quantum state.  Then indeed the
observational probabilities are independent of the quantum state. 
Therefore, there is no logical difficulty in defining probability rules
such that the probabilities of observations are independent of the
quantum state, as is often claimed or wished to be for inflation.

On the other hand, it seems quite {\it ad hoc} to have the observation
operators all be proportional to the identity, so that the observational
probabilities are independent of the quantum state.  If that were the
case, what would be the point of having a quantum state at all in the
theory?  One could just say that the theory consisted of directly giving
the observational probabilities $P_j(i)$ (perhaps from unnormalized
probabilities $p_j(i)$ if they are instrinsically simpler).  If our
observations are indeed independent of the quantum state, why have our
observations been taken to support quantum theory?  That is, why has it
been so successful to unify and simplify the description of our
observations by assuming that they arose from quantum aspects of the
universe, if they come from the expectation values of operators that all
commute?

Therefore, although I have shown here that it is logically possible for
our observations (meaning their probabilities) to have arisen from
probability rules that make them independent of the quantum state, the
way to do this seems highly {\it ad hoc} and implausible.  Surely a
much simpler explanation of our observations will use both a non-random
quantum state and a non-random set of rules for extracting the
probabilities of observations from that quantum state.

\section*{Acknowledgments}

I am grateful for the hospitality of the Perimeter Institute for
Theoretical Physics, where I had long discussions on this subject with
Andreas Albrecht and shorter ones on related issues with Jaume Garriga,
Thomas Hertog, Matthew Kleban, Daniel Phillips, Herman Verlinde, Alex
Vilenkin, and others, though the idea for this paper arose only from
later reflection and was not tested in face-to-face discussions with
any of those scientists.  This research was supported in part by the
Natural Sciences and Engineering Research Council of Canada.

\end{document}